# Magneto-optical characterization of $Mn_xGe_{1-x}$ alloys obtained by ion implantation

F. D'Orazio[a,*], F. Lucari[a], M. Passacantando[a], L. Ottaviano[a], A. Verna[a], G. Impellizzeri[b], F. Priolo[b]

[a]*Dipartimento di Fisica, Università di L'Aquila, I-67010 L'Aquila, Italy*
[b]*IMATIS Lab. and Dipartimento di Fisica e Astronomia, Università di Catania, I-95123 Catania, Italy*

**Abstract**
Magneto-optical Kerr effect hysteresis loops at various wavelengths in the visible/near-infrared range have been used to characterize the magnetic properties of alloys obtained by implanting Mn ions at fixed energy in a Ge matrix. The details of the hysteresis loops reveal the presence of multiple magnetic contributions. They may be attributed to the inhomogeneous distribution of the magnetic atoms and, in particular, to the known coexistence of diluted Mn in the Ge matrix and metallic Mn-rich nanoparticles embedded in it [Phys. Rev. B **73**, 195207 (2006)]

Group-IV semiconductors, doped with magnetic elements, may be a valuable choice for the realization of spintronic devices, due to their potential compatibility with current Si-based processing technology [1,2]. Recently, ion implantation has been revealed a valid alternative to molecular beam epitaxy (MBE) for producing diluted magnetic phases in semiconductors [3,4].

In this work, we report a detailed magneto-optical Kerr effect (MOKE) study of MnGe alloys obtained by implanting $Mn^+$ ions on Ge(100) single crystals. This technique, particularly suitable for characterizing magnetic films, is used as a function of the probing radiation wavelength, for testing the magnetic homogeneity of the material.

During the sample preparation, the Ge wafer was kept at 300 °C while $Mn^+$ ions were implanted with energy equal to 100 keV. The different duration of the process determined the final average of the quasi-gaussian Mn concentration profile inside a layer about 130-140 nm thick [4]. The two samples examined in the present work have doses of $2\times10^{16}$ at./cm$^2$ and $4\times10^{16}$ at./cm$^2$. They will be denoted as samples D2 and D4, respectively. The structural characterization [3,4] reveals the coexistence of diluted Mn in the semiconducting Ge matrix and a fine dispersion of nanoparticles: some constituted by crystalline $Mn_5Ge_3$ with the [0001] axis preferentially perpendicular to the (100) surface of the Ge wafer, others formed by a Mn-rich amorphous GeMn phase. More details on sample preparation and structural characterization are reported in Refs. 3 and 4.

Magnetic analysis, obtained from polar MOKE hysteresis loops, revealed a Curie temperature ($T_C$) of about 255 K and 270 K for the samples D2 and D4, respectively. This difference was related to an enhancement of the long-range order in D4, a feature which manifests also at low temperature with more squared loops and higher coercivity with respect to the sample D2 [4].

In order to investigate more deeply the magneto-optical properties, we collected hysteresis loops at various wavelengths, in the range 0.4 – 2.5 µm, of the s-polarized radiation incident at 45° with respect to the film surface. The maximum amplitude of the magnetic field allowed by the apparatus was 5600 Oe. The measurements were performed at 75 K, since at this temperature the remanence was equal to the zero temperature limit, whereas the irreversible field was smaller than the maximum available for our experimental setup, contrarily to what observed at lower temperatures.

In Fig. 1, we show the MOKE rotation remanence spectra. The behavior is similar for the two samples, except for the different Kerr amplitude, clearly related to the Mn content. As typically expected in MOKE experiments, in polar geometry the signals are larger than in longitudinal geometry. The overall shape of each spectrum is also affected by interference effects related to the finite thickness of the magnetic layer. The comparable position of the peaks and of the zeros for the two samples, since they have similar thickness, is consistent with the reasonable assumption of similar diagonal terms of the dielectric tensor. The attenuation at low wavelength, corresponding to energy much larger than the Ge gap value, is due to loss of transparency of the matrix, when the portion of the sample contributing to the reflected radiation is restricted to the first few layers, characterized by reduced Mn concentration. The remanence spectra obtained from Kerr ellipticity hysteresis loops (not shown) consist, for both samples and for both magnetic field orientations, of a positive curve with a single maximum around 1.6 μm.

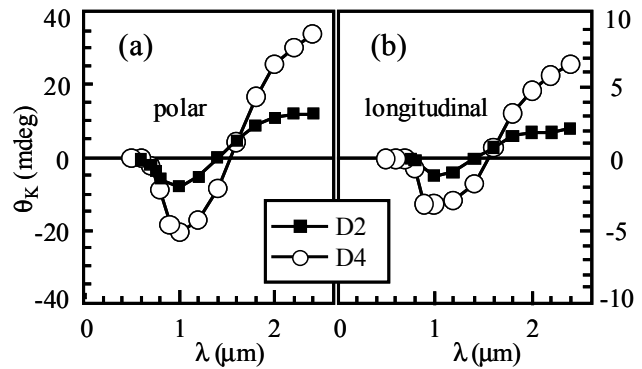

Fig. 1 Polar (a) and longitudinal (b) MOKE rotation remanence spectra at 75 K for the two MnGe alloys.

* Corresponding author. Tel.: +39 0862 433098; fax: +39 0862 433033.
*E-mail address*: franco.dorazio@aquila.infn.it

Additional information concerning the magnetic properties may be achieved by looking in more detail the hysteresis loops. In Figs. 2a and 2b, we show few examples, for both specimens, of polar Kerr rotation at various wavelengths. Longitudinal loops are qualitatively similar to those shown in the figure, although they are generally more squared than polar loops. This may suggest a small degree of shape magnetic anisotropy, which would induce an easier magnetization direction in the film plane. However, due to their larger signals, we restrict our analysis to polar data. For particular wavelengths close to the inversion point of the spectra, the polar hysteresis loops have peculiar shapes. This feature is an indication of the presence of different magnetic phases characterized by different magneto-optical amplitudes and/or signs. For example, the rotation polar loop at 1.6 μm for the sample D2 (Fig. 2c) may only be interpreted as a superposition of two elementary loops, one with positive remanent signal and large hysteresis, the other with negative remanent signal and negligible hysteresis. Clearly, the sign inversion wavelength of the magneto-optical component of each magnetic phase is different: at 1.4 μm (Fig. 2a) the hysteretic loop goes through a zero while the non-hysteretic one is negative. Also at other wavelengths, although less clearly, the presence of more contributions to the hysteresis loops can be recognized. The existence of multiple magnetic phases is confirmed for the sample D4, in particular if we observe the hysteresis loop at 1.6 μm (Fig. 2d). In this case at least three elementary loops must be added in order to fit the experimental curve.

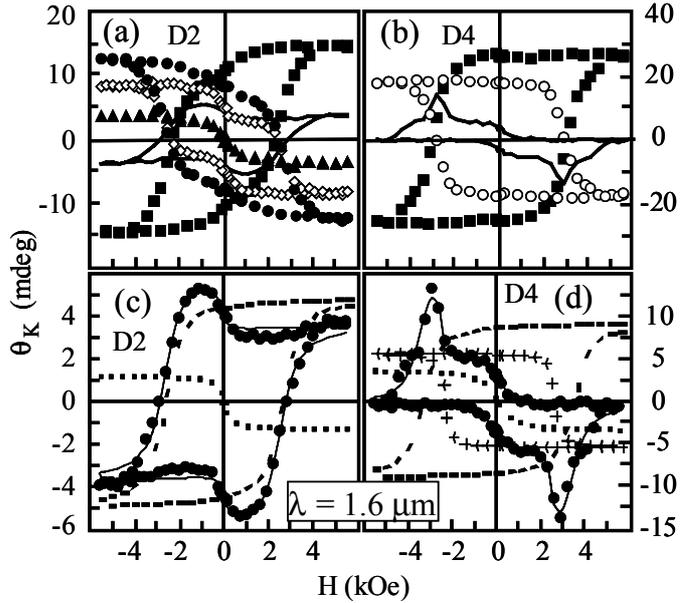

Fig. 2 Polar MOKE hysteresis loops for the samples D2 (a) and D4 (b). In (a) and (b) different symbols indicate different wavelengths (in μm): 2.0 (squares), 1.6 (continuous), 1.4 (triangles), 1.2 (open circles), 1.0 (filled circles). In (c) and (d) the experimental data at 1.6 μm for both samples (black symbols) are fitted by the black line, sum of two (c) or three (d) elementary loops (dashed line, dots, crosses).

The different magneto-optical components must be associated to the inhomogeneous distribution of Mn atoms. We may identify two contributions arising from diluted Ge:Mn and $Mn_5Ge_3$ precipitates, although also the different Mn concentration as a function of depth, originated by the implantation process, may give additional contributions to the overall complex MOKE response.

**References**

[1] Y.D. Park, A.T. Hanbicki, S.C. Erwin, C.S. Hellberg, J.M. Sullivan, J.E. Mattson, T.F. Ambrose, A. Wilson, G. Spanos, and B.T. Jonker, *Science* **295**, 651 (2002).
[2] A.P. Li, J. Shen, J.R. Thompson and H.H. Weitering, *Appl. Phys. Lett*. **86**, 152507 (2005).
[3] L. Ottaviano, M. Passacantando, S. Picozzi, A. Continenza, R. Gunnella, A. Verna, G. Bihlmayer, G. Impellizzeri and F. Priolo, *Appl. Phys. Lett*. **88**, 061907 (2006).
[4] M. Passacantando, L. Ottaviano, F. D'Orazio, F. Lucari, M. De Biase, G. Impellizzeri, F. Priolo, *Phys. Rev. B* **73**, 195207 (2006).